# Inverse Design of an On-chip Phase-Change Switch Composed of Sparse Heterogeneous Metasurfaces

## Sanaz Zarei

Sharif University of Technology, Tehran, Iran
szarei@sharif.edu

***Abstract*—Photonic switches are exploited for optical signal routing and control within optical communication systems. The integration of phase change materials with silicon photonic switches overcomes the drawbacks of the state-of-the-art photonic switches based on electro-optic and thermo-optic effects. In this work, a 1×2 on-chip switch with a footprint of 6.3μm×31μm based on on-chip one-dimensional metasurfaces is inverse-designed. The on-chip metasurfaces are heterogeneous, which their meta-atoms can be composed of either $SiO_2$ or the phase-change material $Sb_2Se_3$. This helps to the realization of higher extinction ratio between the switch outputs, especially at the amorphous state which has a small refractive index contrast to that of the silicon. To benefit from structural, fabrication and switching simplicity, sparse parameterization is followed in the inverse design of the switch using ternary search (TS) and particle swarm optimization (PSO) algorithms.**

## 1. Introduction

Silicon photonic switches are fundamental to optical communication networks [1]. They are also the pivotal building blocks of programmable integrated photonic circuits (PICs) [2]. The current photonic switches are mainly based on Mach-Zehnder interferometers or micro ring resonators working with thermo-optic or electro-optic effects [1-2] and suffer from a large programming energy, small change in their refractive index, large footprint, and a volatile nature [3]. Hence, a switch with compact footprint, low energy consumption, low insertion loss (IL) and high extinction ratios (ER) is highly demanded [1-2]. Recently, phase change materials (PCMs) have attracted increasing attention for realizing non-volatile photonic switches. Among PCMs, antimony tri-selenide ($Sb_2Se_3$) has emerged as a promising PCM due to its extremely low intrinsic absorption losses at both amorphous and crystalline phases and a large refractive index contrast of $\Delta n = 0.77$ between the two states over the telecommunication transmission band [3].

Here, a $Sb_2Se_3$-enabled on-chip switch based on one-dimensional (1D) on-chip metasurfaces [3-8] is inverse-designed using ternary search (TS) and particle swarm optimization (PSO) algorithms. The infrastructural on-chip 1D metasurfaces are sparse and heterogeneous. While a heterogeneous metasurface is an engineered planar structure composed of multiple distinct materials or varied nanostructured geometries (to achieve complex wavefront and dispersion control), a sparse metasurface is a nonperiodic engineered structure that uses a low-density arrangement of scattered elements (meta-atoms) rather than a dense, continuous grid (to simplify design and manufacturing process and reduce loss). In the presented scheme, instead of $Sb_2Se_3$, $SiO_2$ is used for material filling of some meta-atoms in the 1D metasurfaces, which helps to improve the extinction ratios of the switch, especially at the amorphous state [1]. Furthermore, the unnecessary ($Sb_2Se_3$ or $SiO_2$) cells that have minimal impact on the performance, are dropped during the switch optimization, to simplify the design.

## 2. Phase-change Switch Design

The first design phase is accomplished using a ternary search algorithm, for which its flowchart is illustrated in Fig. 1(a). For all optimizations, the 2.5D variational FDTD solver of Lumerical Mode Solution (varFDTD) is used for structural modeling. The optimization is performed in a metasystem consisting of five 1D metasurfaces (i.e. metalines) with 6μm separations. The first metaline is located at x=0 and the output line with two output monitors is fixed at x = 30μm. The output monitors are 0.9μm wide and 3μm center-to-center distant. For excitation, a mode source with 0.9μm width, x-axis as the injection axis, and 1.55μm as the excitation wavelength, is placed at x=-1μm, with its center aligned exactly with the center of the metalines. Each metaline initially has 21 meta-atoms, each of which consists of an etched rectangular slot (with a slot width of 100nm, length of 2μm, and height of 220nm) in the silicon-on-insulator (SOI) substrate, that is filled with $Sb_2Se_3$ [9]. The thickness of the silicon device layer and the $SiO_2$ insulator layer are 220nm and 2μm, respectively. There is a 2μm-thick $SiO_2$ topcoat on the silicon layer for protection and diffusion inhibition of $Sb_2Se_3$. The pitch of metalines is set to be 300nm. To realize heterogeneity and sparsity, throughout the TS algorithm, each slot within the metalines, can be filled with Si, $Sb_2Se_3$ or $SiO_2$. By Si filling, the unnecessary ($Sb_2Se_3$ or $SiO_2$-filled) slots will be withdrawn, which simultaneously decreases the number of phase-change ($Sb_2Se_3$) cells and $SiO_2$ rods, ease the fabrication process, and maximize the performance. A figure of merit (FOM) is specified to evaluate the optimized performance. The material within the slots is serially switched in a manner like $Sb_2Se_3 \rightarrow SiO_2 \rightarrow Si \rightarrow Sb_2Se_3$ (see Fig. 1(a)) and for each change, the FOM is calculated for both amorphous and crystalline states. Each meta-atom captures the material with the least FOM and the algorithm switches to the next meta-atom. The defined FOM is as below:

$$FOM = FOM_{amorphous} + FOM_{crystalline}, \quad FOM_{amorphous} = |1 - T_1| + |T_2|, \quad FOM_{crystalline} = |T_1| + |1 - T_2| \tag{1}$$

where $T_1$ is the transmittance of the top monitor and $T_2$ is the transmittance of the bottom monitor (see the inset of Fig. 1(b)). The specified FOM aims to maximize $T_1$ and minimize $T_2$ in the amorphous state, while minimizing $T_1$ and maximizing $T_2$ in the crystalline state. Figure 1(b) shows the FOM variation during the TS optimization. The FOM is initially equal to 2 and as the optimization continues, it reduces to 1.1094 at the nineth iteration and no further change happens afterwards. The optimization is terminated after the twelfth iteration. The inset of Fig. 1(b) presents the optimized metasystem and Fig. 1(c) exhibits the optimized transmissions of the output monitors at both states, calculated by varFDTD and verified by 3D FDTD (Lumerical FDTD). Figures 1(d) and 1(e) display the x-y view of the electric field distribution at the midplane of the silicon layer in the optimized metasystem, at amorphous and crystalline states, respectively.

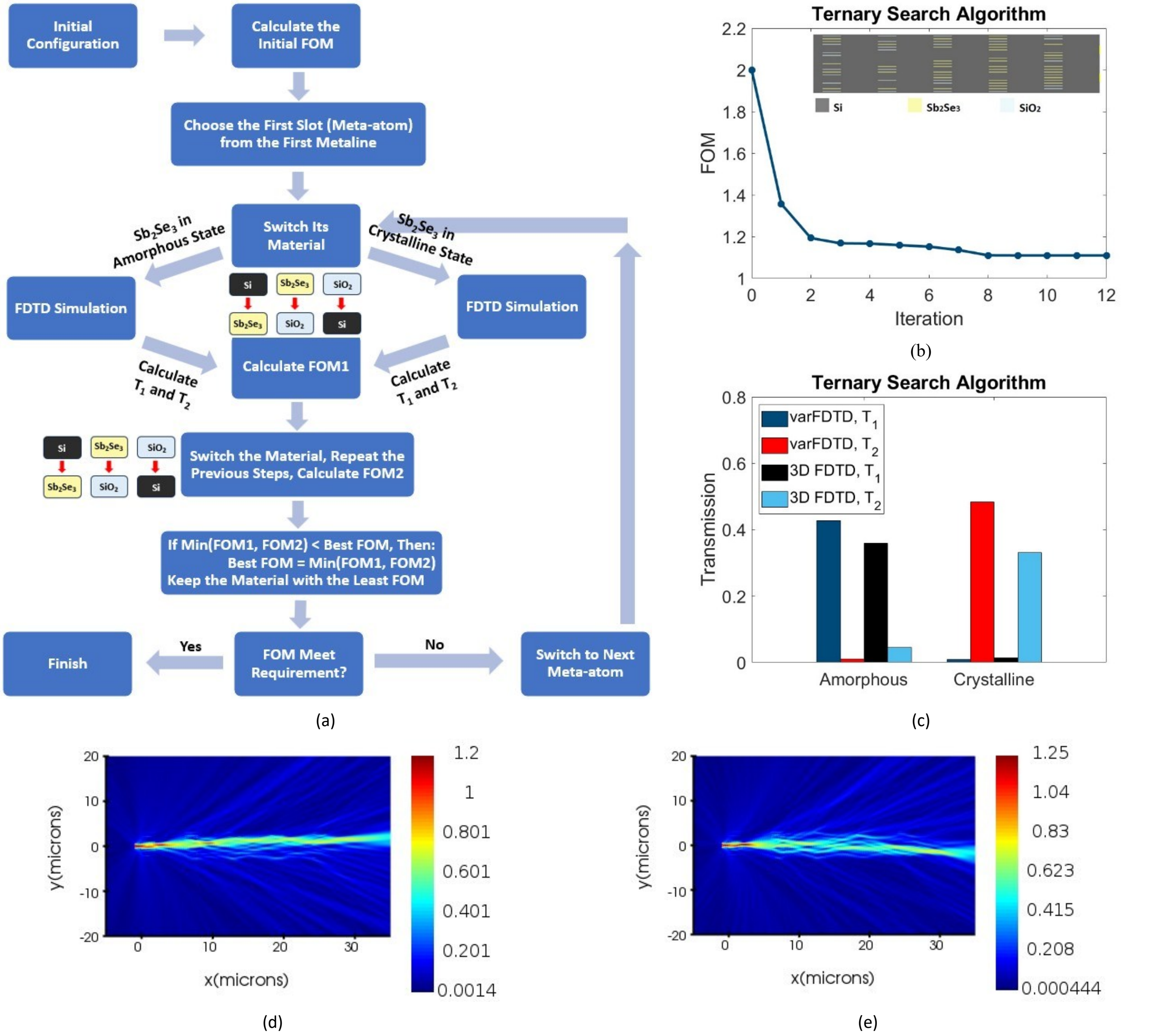

Fig. 1. (a) The flowchart of the ternary search algorithm, (b) The FOM change versus iteration throughout the TS optimization; the inset represents the optimized metasystem. (c) The output (absolute) transmissions in the optimized metasystem at both amorphous and crystalline sates, calculated by varFDTD and verified by 3D FDTD. The x-y view of the electric field distribution at the midplane of the silicon device layer in the optimized metasystem, (d) in amorphous state, (e) in crystalline state.

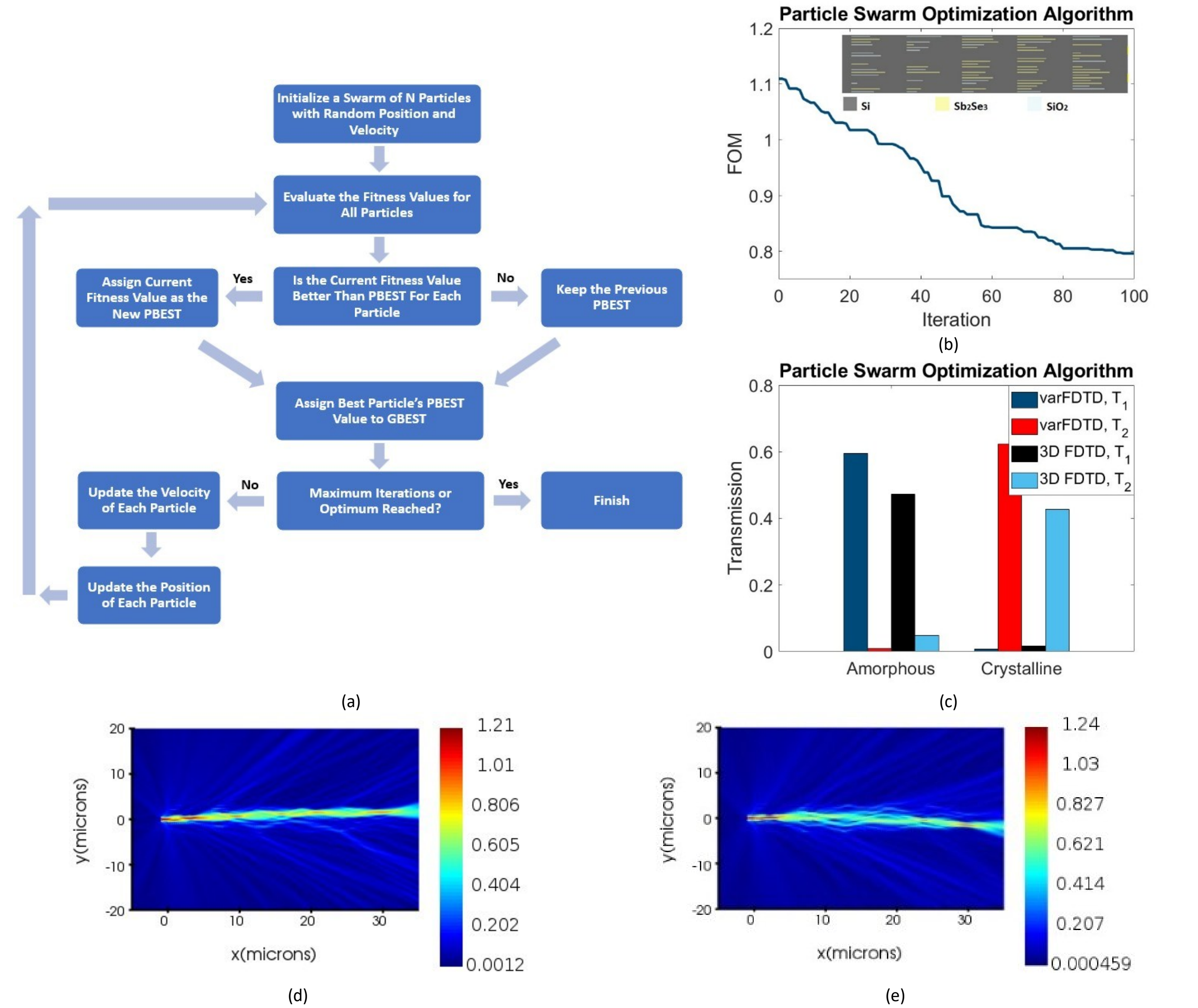

Fig. 2. (a) The flowchart of the particle swarm optimization algorithm, (b) The FOM change versus iteration throughout the PSO optimization; the inset shows the optimized metasystem. (c) The output (absolute) transmissions in the optimized metasystem at both amorphous and crystalline sates, calculated by varFDTD and verified by 3D FDTD. The x-y view of the electric field distribution at the midplane of the silicon device layer in the optimized metasystem, (d) in amorphous state, (e) in crystalline state.

The optimized metasystem by TS algorithm is additionally optimized by a particle swarm optimization (PSO) algorithm in the second design phase. The flowchart of the PSO optimization is depicted in Fig. 2(a). In this phase, the lengths of all slots that are filled with either $Sb_2Se_3$ or $SiO_2$ are adjusted within the range of 100nm and 4.5µm (but the width and height of the slots remain unchanged) to further minimize the FOM. As can be observed in Fig. 2(b), the FOM is minimized to 0.7961 in 100 iterations. The inset shows the optimized metasystem. Fig. 2(c) elucidates the transmissions of the output monitors in the optimized metasystem at both amorphous and crystalline states, calculated by both varFDTD and 3D FDTD. The x-y view of the electric field distribution at the middle plane of the silicon layer within the optimized structure for amorphous and crystalline states, is shown in figures 2(d) and 2(e), respectively.

So far, the optimizations have been performed within a metasystem composed of five cascaded 1D metasurfaces and the consideration of $SiO_2$ as an alternative filling material for slots leads to almost equal extinction ratio between the outputs at both amorphous and crystalline states (comparing to a metasystem with only $Sb_2Se_3$ filling [9]), as can be seen in figures 1(c) and 2(c).

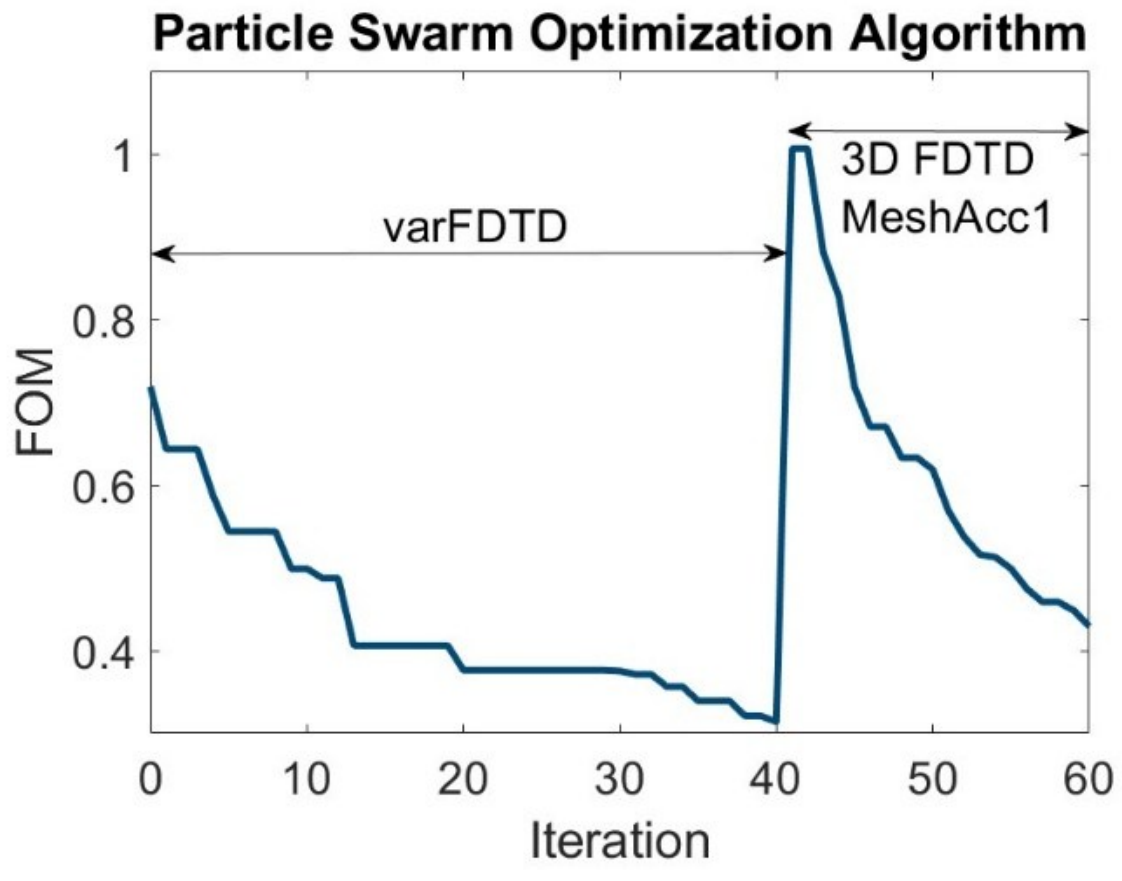

(a)

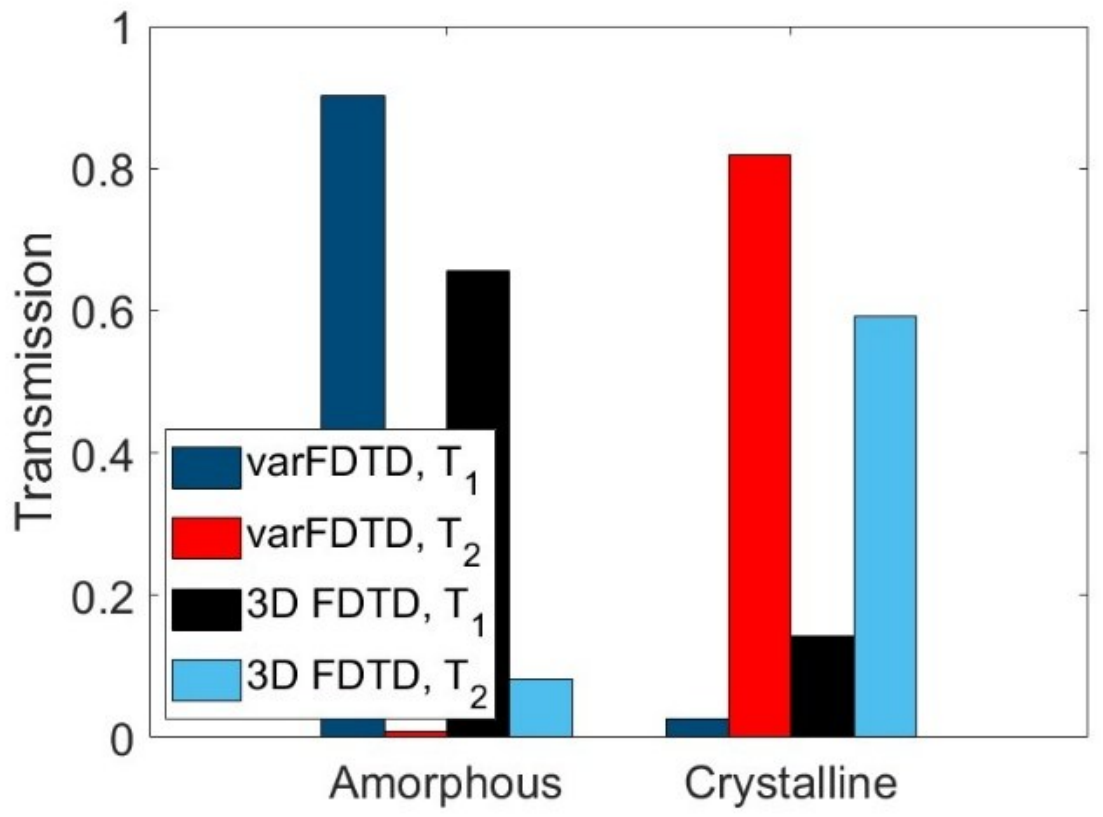

(b)

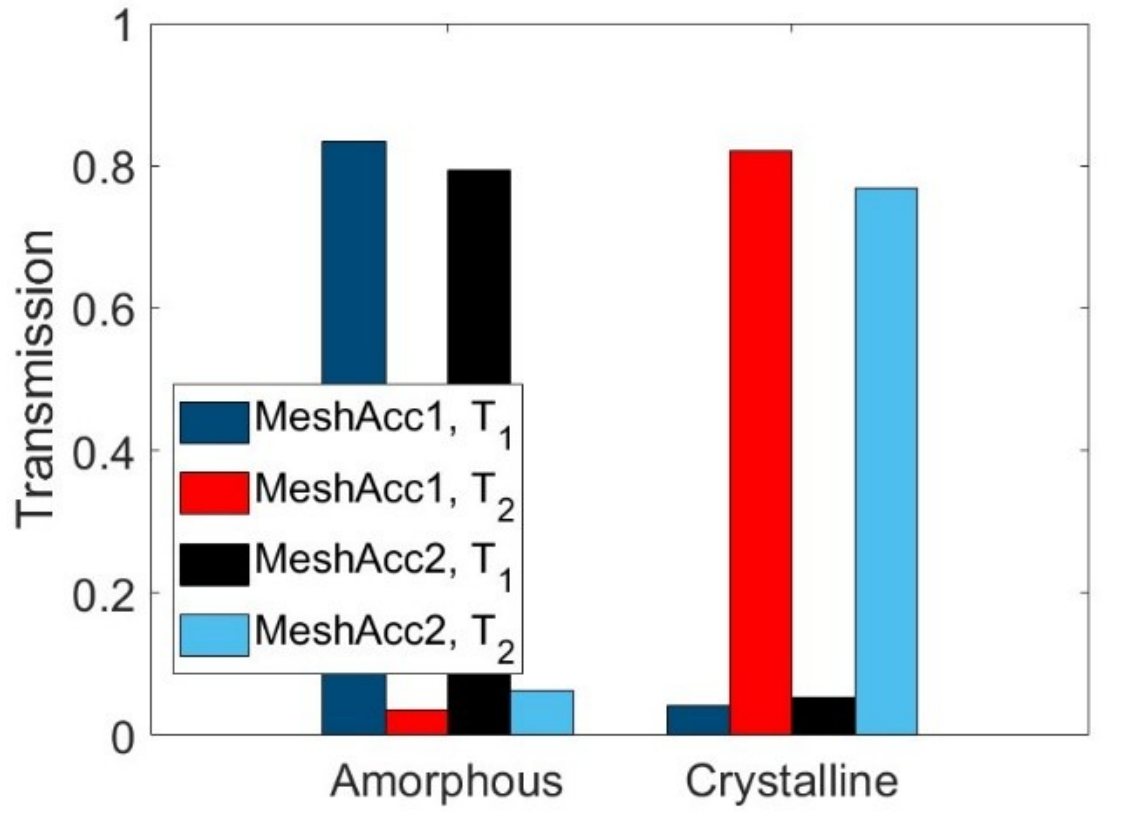

(c)

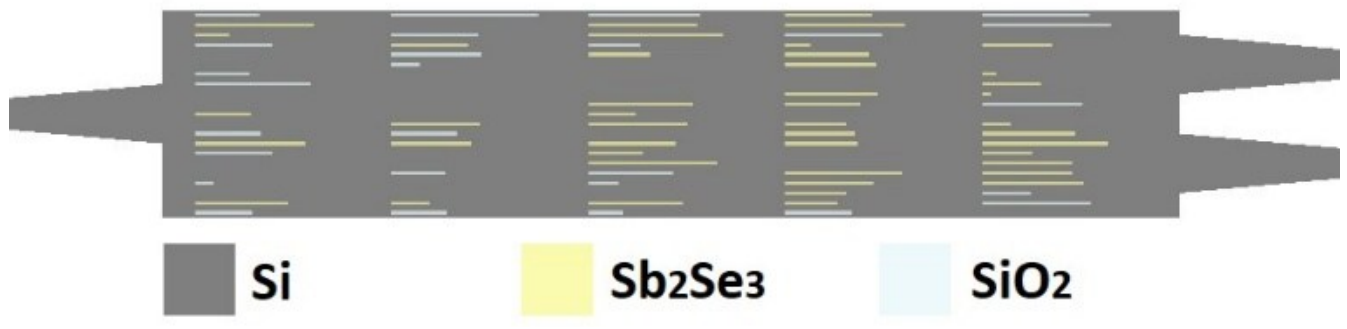

(d)

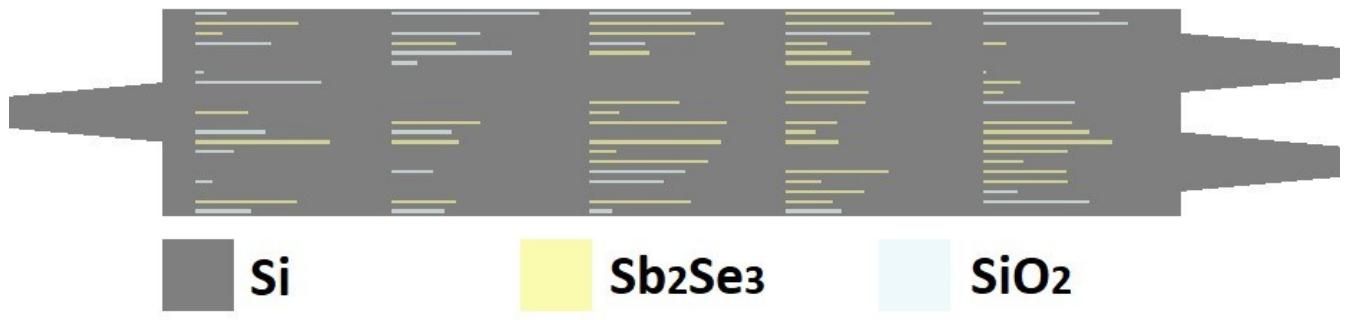

(e)

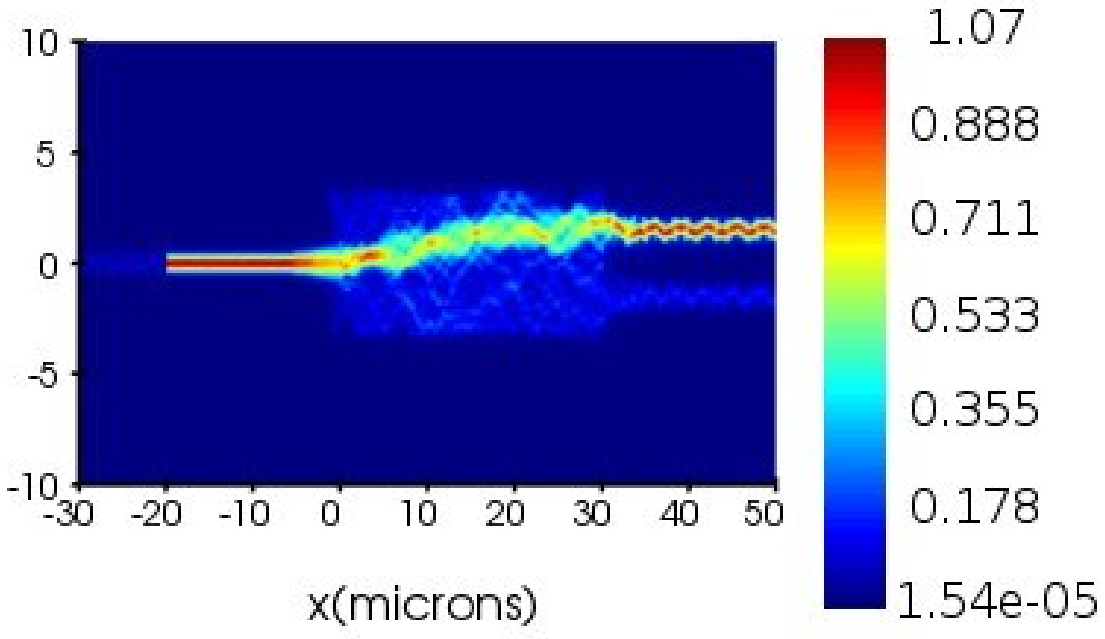

(f)

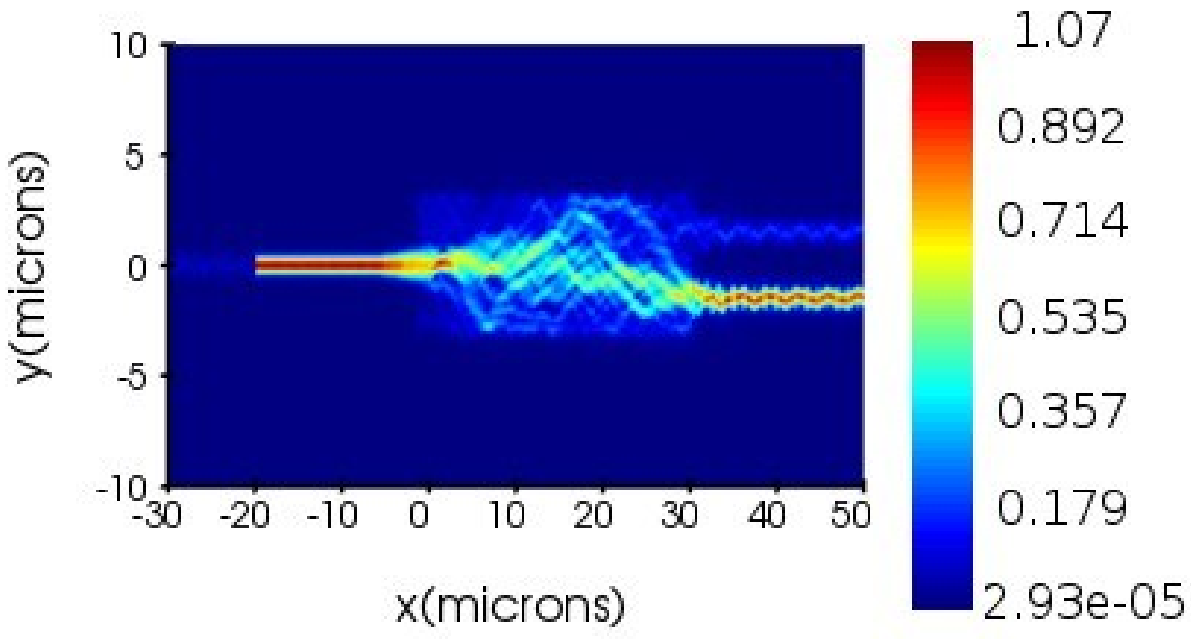

(g)

Fig. 3. (a) The FOM change versus iteration throughout the final PSO optimization, (b) The output (absolute) transmissions of the optimized metasystem at both amorphous and crystalline states, after 40 iterations of optimization using varFDTD for structural modeling, verified by 3D FDTD. (c) The output (absolute) transmissions of the optimized metasystem at both amorphous and crystalline states, after the last 20 iterations of optimization using 3D FDTD (with mesh accuracy setting to 1) for structural modeling, verified by 3D FDTD with the higher mesh accuracy of 2. (d) Schematic of the optimized 1×2 on-chip switch after 40 iterations. (e) Schematic of the finally-optimized 1×2 on-chip switch. The x-y view of the electric field distribution at the midplane of the silicon device layer in the optimized metasystem (f) in amorphous state, (g) in crystalline state.

It is worth mentioning that for these metasystems, the discrepancy between varFDTD and 3D FDTD results is larger, comparing to the metasystems comprising of only $Sb_2Se_3$ cells [9], that can be attributed to the existence of $SiO_2$ rods in the metasystem with a relatively large refractive index contrast to that of silicon.

To propose a switch scheme, the finally optimized metasystem is transferred into a 1×2 configuration like the ones shown in figures 3(d) and 3(e). Such a configuration with the main body footprint of 6.3µm×31µm, comprises input/output waveguides of 0.9µm width and input/output tapers of 1.8µm width (in the place of conjunction with the main body) and 5µm length. The center-to-center distance between the output waveguides/tapers is 3µm. By this transfer, the FOM reduces to 0.7189 and the electric field distribution at the output line (and within the output monitors) changes moderately. Further optimization is thus required to finalize the design. Therefore, a second PSO optimization is applied to optimize the new metasystem and improve the performance. This third optimization phase lasts for 60 iterations. The first 40 iterations are accomplished using varFDTD for structural modeling. After this optimization, the optimized output transmissions in Fig. 3(b) are achieved at both amorphous and crystalline states. The optimized metasystem is pictured in Fig. 3(d). Because of the large discrepancy between varFDTD results and 3D FDTD verifications, which leads to the degraded performance of the switch design, the last 20 iterations are performed using 3D FDTD for structural modeling. To achieve feasible simulation and optimization times, the mesh accuracy is set to 1 within the Lumerical FDTD software. The verified results using mesh accuracy 2 has small differences with optimized ones, which is shown in Fig. 3(c). The finally optimized switch is presented in Fig. 3(e). Figures 3(f) and 3(g) exhibit the x-y view of the electric field distribution in the optimized metasystem at both amorphous and crystalline states.

The insertion loss of the optimized switch at both amorphous and crystalline states, calculated from equation (2), are 0.94dB and 1.1dB, respectively. In equation (2), $T_{in}$ is the transmittance from the input taper/waveguide to the switch body. The computed extinction ratio, based on equation (3), is approximately 11dB at amorphous state and 11.6dB at crystalline state.

$$IL = -10\log(T_{1(2)} / T_{in}) \tag{2}$$

$$ER = 10\log(T_{1(2)} / T_{2(1)}) \tag{3}$$

Although heterogeneous on-chip metasurfaces helps to achieve better extinction ratio at both amorphous and crystalline states in most of the accomplished optimizations (figures 1(c), 2(c) and 3(b)), the presented design with 44 $Sb_2Se_3$ phase-change cells and 27 $SiO_2$ rods, fails to precede the previous design that only consists of $Sb_2Se_3$ rods [9]. As is evident in figures 1(c), 2(c) and 3(b), this is eventuated by the limitations of the varFDTD that is used for structural modeling. 3D FDTD is definitely a better choice for performing the optimizations, however, the simulation time is a big barrier. More investigation is required to overcome the obstructions.

## 3. Conclusion

In summary, a nonvolatile and compact 1×2 on-chip switch with a footprint of 6.3µm×31µm is demonstrated using on-chip heterogeneous metasurfaces. The heterogeneous metasurfaces are sparse metasurfaces with meta-atoms that are composed of either $SiO_2$ or the phase-change material $Sb_2Se_3$. This scheme aims to the realization of lower loss and more easily fabricable and switchable device with minimum number of phase-change ($Sb_2Se_3$) and $SiO_2$ cells that realizes higher extinction ratio between the switch outputs especially at the amorphous state which has a small refractive index contrast to that of the silicon. The presented switch exhibits an insertion loss of less than 1.1dB and extinction ratio between the outputs of more than 11dB.

## Statements & Declarations

### Funding

The author declares that no funds, grants, or other support were received during the preparation of this manuscript.

### Competing Interests

The author has no relevant financial or non-financial interests to disclose.

### Author Contributions

S. Z. performed the study conception and design, accomplished the FDTD simulations, carried out the TS and PSO optimizations, analyzed the results, and wrote the manuscript.

### Data Availability

No datasets were generated or analyzed during this study.

### Ethics declarations

This article does not contain any studies involving animals or human participants.